\documentclass[prb,twocolumn,showpacs]{revtex4}
\usepackage{amsmath}
\usepackage{color,array,dcolumn}
\usepackage{dcolumn}
\usepackage{graphicx}
\usepackage{bm}

\begin{document}

\title{Weak ferromagnetism with the Kondo screening effect in the Kondo
lattice systems}
\author{Yu Liu$^{1}$, Guang-Ming Zhang$^{1}$, and Lu Yu$^{2}$}
\affiliation{$^{1}$State Key Laboratory of Low-Dimensional Quantum Physics and Department
of Physics, Tsinghua University, Beijing 100084, China\\
$^{2}$Beijing National Laboratory for Condensed Matter Physics and\\
Institute of Physics, Chinese Academy of Sciences, Beijing 100190, China}
\date{\today }

\begin{abstract}
We carefully consider the interplay between ferromagnetism and the Kondo
screening effect in the conventional Kondo lattice systems at finite
temperatures. Within an effective mean-field theory for small conduction
electron densities, a complete phase diagram has been determined. In the
ferromagnetic ordered phase, there is a characteristic temperature scale to
indicate the presence of the Kondo screening effect. We further find two
distinct ferromagnetic long-range ordered phases coexisting with the Kondo
screening effect: spin fully polarized and partially polarized states. A
continuous phase transition exists to separate the partially polarized
ferromagnetic ordered phase from the paramagnetic heavy Fermi liquid phase.
These results may be used to explain the weak ferromagnetism observed
recently in the Kondo lattice materials.
\end{abstract}

\pacs{71.10.Fd, 71.27.+a, 71.30.+h, 75.20.Hr}
\maketitle

The most important issue in the study of heavy fermion materials is the
interplay between the Kondo screening and the magnetic interactions among
local magnetic moments mediated by the conduction electrons.\cite%
{Stewart-2001,Lohneysen-2007,Steglich-2010} The former effect favors the
formation of Kondo singlet state in the strong Kondo coupling limit, while
the latter interactions tend to stabilize a magnetically long-range ordered
state in the weak Kondo coupling limit. In-between these two distinct
phases, there exists a magnetic phase transition. Although such a phase
transition was suggested by Doniach many years ago,\cite{Doniach1977,
Lacroix1979} the complete finite temperature phase diagram for the Kondo
lattice systems has not been derived from a microscopic theory.\cite{Q-Si}
At the half-filling of the conduction electrons, the antiferromagnetic
long-range order dominates over the local magnetic moments, which can be
partially screened by the conduction electrons in the intermediate Kondo
coupling regime.\cite{Zhang2000a,Assaad2001,Ogata2007,Assaad-2008} Very
recently, close to the magnetic phase transition, weak ferromagnetism below
the Kondo temperature has been discovered in the Kondo lattice materials UCu$%
_{5-x}$Pd$_{x}$ (Ref.\onlinecite{Bernal-1995}), URh$_{1-x}$Ru$_{x}$Ge (Ref.%
\onlinecite{Huy-2007}),YbNi$_{4}$P$_{2}$ (Ref.\onlinecite{Krellner-2011}),
YbCu$_{2}$Si$_{2}$ (Ref.\onlinecite{Flouquet-2011}), and Yb(Rh$_{0.73}$Co$%
_{0.27}$)$_{2}$Si$_{2}$ (Ref.\onlinecite{Lausberg-2012}). So an interesting
question arises as whether the ferromagnetic long-range order can coexist
with the Kondo screening effect.

To account for the ferromagnetism within the Kondo lattice model, one can
assume that conduction electrons per local moment $n_{c}$ is far away from
half-filling, where the ferromagnetic correlations dominate in the small
Kondo coupling regime.\cite{IK-1991,Sigrist-1992,Li-1996,Si} Similar to the
interplay between the antiferromagnetic correlations and the Kondo screening
effect argued by Doniach,\cite{Doniach1977} a schematic finite temperature
phase diagram can be argued for the interplay between the ferromagnetic
correlations and the Kondo screening effect. In Fig.1, the Curie temperature
is plotted as a function of the Kondo coupling. For the small Kondo
couplings, the ferromagnetic ordering (Curie) temperature is larger than the
single-impurity Kondo temperature. When the Kondo coupling is strong enough,
the Curie temperature is suppressed completely. However, there is an
important issue as whether there should be a characteristic temperature
scale inside the ferromagnetic ordered phase to signal the presence of the
Kondo screening effect. If so, there may exist two distinct ferromagnetic
ordered phases: a pure ferromagnetic phase with a small Fermi surface
consisting of conduction electrons only, and a ferromagnetic phase with an
enlarged Fermi surface including both conduction electrons and local
magnetic moments, coexisting with the Kondo screening.
\begin{figure}[tbp]
\includegraphics[scale=0.5]{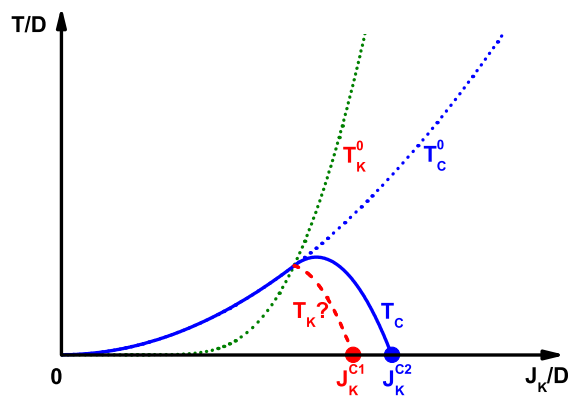}
\caption{(Color online) The schematic phase diagram expected from the
interplay between ferromagnetic correlations and the Kondo screening effect.
$T_{C}^{0}$ denotes the Curie temperature in the absence of the Kondo
effect, and $T_{K}^{0}$ represents the Kondo temperature without the
ferromagnetic correlations.}
\end{figure}

In our previous paper,\cite{Li-Zhang-Yu} we have carefully studied the
possible ground state phases within an effective mean field theory. In
particular, for $0.16<n_{c}<0.82$ and close to the magnetic phase
transition, the local moments can be only partially screened by the
conduction electrons, and the remaining uncompensated parts develop the
ferromagnetic long-range order. Depending on the Kondo coupling strength,
the resulting ground state is either a spin fully polarized or a partially
polarized ferromagnetic phase according to the quasiparticles around the
Fermi energy. The existence of the spin fully polarized coexistent Kondo
ferromagnetic phase has been confirmed by the recent dynamic mean-field
calculations in infinite dimensions and density matrix renormalization group
in one dimension, where such a state is referred to as the spin-selective
Kondo insulator.\cite{Peters-2012}

In the present paper, we will derive a similar finite temperature
phase diagram of the Kondo lattice model to Fig.1 for small
conduction electron densities. Below the Curie temperature, we
find a characteristic temperature scale to signal the Kondo
screening effect for the first time. Moreover, there exist a spin
fully polarized phase and a partially polarized ferromagnetic
long-range ordered phase coexisting with the Kondo screening
effect. The former phase spans a large area in the phase diagram,
while the latter phase just occupies a very narrow region close to
the phase boundary of the paramagnetic heavy Fermi liquid phase.
Moreover, a second order phase transition occurs from the spin
partially polarized ferromagnetic ordered state to the
paramagnetic heavy Fermi liquid state, and the transition line
becomes very steep close to the quantum critical point. Our
results may be used to explain the weak ferromagnetism and quantum
critical behavior observed in
YbNi$_{4}$P$_{2}$.\cite{Krellner-2011}

The Hamiltonian of the Kondo lattice systems is defined by%
\begin{equation}
\mathcal{H}=\sum_{\mathbf{k},{\sigma }}\epsilon _{\mathbf{k}}c_{\mathbf{k}%
\sigma }^{\dagger }c_{\mathbf{k}\sigma }+J_{K}\sum_{i}\mathbf{\sigma }%
_{i}\cdot \mathbf{S}_{i},
\end{equation}%
where $\epsilon _{\mathbf{k}}$ is the dispersion of the conduction
electrons, $\mathbf{\sigma }_{i}=\frac{1}{2}\sum_{{\alpha }\beta }c_{i\alpha
}^{\dagger }\mathbf{\tau }_{\alpha \beta }c_{i\beta }$ is the spin density
operator of the conduction electrons, $\mathbf{\tau }$ is the Pauli matrix,
and the Kondo coupling strength $J_{K}>0$. When the localized spins are
denoted by $\mathbf{S}_{i}=\frac{1}{2}\sum_{{\alpha }\beta }f_{i\alpha
}^{\dagger }\mathbf{\tau }_{\alpha \beta }f_{i\beta }$ in the pseudo-fermion
representation, the projection onto the physical subspace has to be
implemented by a local constraint $\sum_{\sigma }f_{i\sigma }^{\dagger
}f_{i\sigma }=1$. It is straightforward to decompose the Kondo spin exchange
into longitudinal and transversal parts
\begin{equation*}
\mathbf{\sigma }_{i}\cdot \mathbf{S}_{i}=\sigma _{i}^{z}S_{i}^{z}-\frac{1}{4}%
[(c_{i\uparrow }^{\dagger }f_{i\uparrow }+f_{i\downarrow }^{\dagger
}c_{i\downarrow })^{2}+(c_{i\downarrow }^{\dagger }f_{i\downarrow
}+f_{i\uparrow }^{\dagger }c_{i\uparrow })^{2}],
\end{equation*}%
where the longitudinal part describes the polarization of the conduction
electrons, giving rise to the usual RKKY interaction among the local
moments; while the transverse part represents the spin-flip scattering of
the conduction electrons by the local moments, yielding the local Kondo
screening effect.\cite{Lacroix1979,Zhang2000a} The latter effect has been
investigated by various approaches, in particular, those based on a $1/N$
expansion\cite{read,coleman,burdin} ($N$ is the degeneracy of the localized
spin). However, the competition between these two interactions determines
the possible ground states of the Kondo lattice systems.

Let us first review the effective mean field theory for the ground state
used in our previous study.\cite{Li-Zhang-Yu} We introduce two ferromagnetic
order parameters: $m_{f}=\left\langle S_{i}^{z}\right\rangle $ and $%
m_{c}=\langle \sigma _{i}^{z}\rangle $ to decouple the longitudinal exchange
term, while a hybridization order parameter $V=\langle c_{i\uparrow
}^{\dagger }f_{i\uparrow }+f_{i\downarrow }^{\dagger }c_{i\downarrow
}\rangle $ is introduced to decouple the transverse exchange term. We also
introduce a Lagrangian multiplier $\lambda $ to enforce the local
constraint, which becomes the chemical potential in the mean field
approximation. Then the mean field Hamiltonian in the momentum space can be
written in a compact form
\begin{equation}
\mathcal{H}_{MF}=\sum_{\mathbf{k},{\sigma }}\left( c_{\mathbf{k}\sigma
}^{\dagger },f_{\mathbf{k}\sigma }^{\dagger }\right) \left(
\begin{array}{cc}
\epsilon _{\mathbf{k}{\sigma }} & -\frac{J_{K}V}{2} \\
-\frac{J_{K}V}{2} & \lambda _{{\sigma }}%
\end{array}%
\right) \left(
\begin{array}{c}
c_{\mathbf{k}\sigma } \\
f_{\mathbf{k}\sigma }%
\end{array}%
\right) +\mathcal{N}\varepsilon _{0},
\end{equation}%
where $\epsilon _{\mathbf{k}{\sigma }}=\epsilon _{\mathbf{k}}+\frac{%
J_{K}m_{f}}{2}{\sigma }$, $\lambda _{{\sigma }}=\lambda +\frac{J_{K}m_{c}}{2}%
{\sigma }$, $\varepsilon _{0}=\frac{J_{K}V^{2}}{2}-J_{K}m_{c}m_{f}-\lambda $%
, ${\sigma =\pm 1}$ denote the up and down spin orientations, and $\mathcal{N%
}$ is the total number of lattice sites. The quasiparticle excitation
spectra are thus obtained by
\begin{equation}
E_{\mathbf{k}{\sigma }}^{\pm }=\frac{1}{2}\left[ \epsilon _{\mathbf{k}{%
\sigma }}+\lambda _{{\sigma }}\pm \sqrt{\left( \epsilon _{\mathbf{k}{\sigma }%
}-\lambda _{{\sigma }}\right) ^{2}+(J_{K}V)^{2}}\right] ,
\end{equation}%
where there appear four quasiparticle bands with spin splitting.

Using the method of equation of motion, the single particle Green functions
can be derived, while the corresponding density of states can be calculated
and expressed as
\begin{gather}
\rho _{c}^{\sigma }(\omega )=\rho _{c}^{0}[\theta (\omega -\omega _{1\sigma
})\theta (\omega _{2\sigma }-\omega )+\theta (\omega -\omega _{3\sigma
})\theta (\omega _{4\sigma }-\omega )],  \notag \\
\rho _{f}^{\sigma }(\omega )=\left( \frac{J_{K}V/2}{\omega -\lambda _{{%
\sigma }}}\right) ^{2}\rho _{c}^{\sigma }(\omega ),
\end{gather}%
where $\theta (\omega )$ is a step function and a constant density of states
of conduction electrons has been assumed $\rho _{c}^{0}=\frac{1}{2D}$, with $%
D$ as a half-width of the conduction electron band. The four quasiparticle
band edges can be expressed as
\begin{align*}
\omega _{1\sigma }& =\frac{1}{2}\left[ \epsilon _{\sigma }-D+\lambda _{{%
\sigma }}-\sqrt{(\epsilon _{\sigma }-D-\lambda _{{\sigma }})^{2}+(J_{K}V)^{2}%
}\right] , \\
\omega _{2\sigma }& =\frac{1}{2}\left[ \epsilon _{\sigma }+D+\lambda _{{%
\sigma }}-\sqrt{(\epsilon _{\sigma }+D-\lambda _{{\sigma }})^{2}+(J_{K}V)^{2}%
}\right] , \\
\omega _{3\sigma }& =\frac{1}{2}\left[ \epsilon _{\sigma }-D+\lambda _{{%
\sigma }}+\sqrt{(\epsilon _{\sigma }-D-\lambda _{{\sigma }})^{2}+(J_{K}V)^{2}%
}\right] , \\
\omega _{4\sigma }& =\frac{1}{2}\left[ \epsilon _{\sigma }+D+\lambda _{{%
\sigma }}+\sqrt{(\epsilon _{\sigma }+D-\lambda _{{\sigma }})^{2}+(J_{K}V)^{2}%
}\right] ,
\end{align*}%
where $\epsilon _{\sigma }=\frac{J_{K}m_{f}}{2}{\sigma }$ and $\omega
_{1\sigma }<\omega _{2\sigma }<\omega _{3\sigma }<\omega _{4\sigma }$.

Then using the spectral representation of the Green functions, we derive the
mean-field equations at finite temperatures as follows
\begin{eqnarray}
\int_{-\infty }^{+\infty }d\omega f(\omega )\left[ \rho _{c}^{+}(\omega
)+\rho _{c}^{-}(\omega )\right] &=&n_{c},  \notag \\
\int_{-\infty }^{+\infty }d\omega f(\omega )\left[ \rho _{c}^{+}(\omega
)-\rho _{c}^{-}(\omega )\right] &=&2m_{c},  \notag \\
\sum\limits_{\sigma }\int_{-\infty }^{+\infty }d\omega f(\omega )\frac{\rho
_{c}^{\sigma }(\omega )}{\left( \lambda _{\sigma }-\omega \right) ^{2}}%
\left( \frac{J_{K}V}{2}\right) ^{2} &=&1,  \notag \\
\sum\limits_{\sigma }\int_{-\infty }^{+\infty }d\omega f(\omega )\frac{%
\sigma \rho _{c}^{\sigma }(\omega )}{\left( \lambda _{\sigma }-\omega
\right) ^{2}}\left( \frac{J_{K}V}{2}\right) ^{2} &=&2m_{f},  \notag \\
\sum\limits_{\sigma }\int_{-\infty }^{+\infty }d\omega f(\omega )\frac{\rho
_{c}^{\sigma }(\omega )}{\left( \lambda _{\sigma }-\omega \right) }\left(
\frac{J_{K}V}{2}\right) &=&V,
\end{eqnarray}%
where $f(\omega )=1/\left[ 1+e^{(\omega -\mu )/T}\right] $ is the Fermi
distribution function. To make the magnetic interaction between the nearest
neighboring local moments ferromagnetic, we should confine the density of
conduction electrons to $n_{c}<0.82$ from the previous mean field study.\cite%
{Fazekas1991}

The position of the chemical potential $\mu $ with respect to the band edges
is very important. At zero temperature, there are two different situations.
The corresponding schematic local density of states are displayed in Fig.2.
For $\omega _{1-}<\mu <\omega _{2+}$, both the lower spin-up and spin-down
quasiparticle bands are partially occupied, corresponding to the spin
partially polarized ferromagnetic state. However, for $\omega _{2+}<\mu
<\omega _{2-}$, the lower spin-up quasiparticle band is completely occupied,
while the lower spin-down quasiparticle band is only partially occupied,
corresponding to the spin fully polarized ferromagnetic state.\cite%
{Beach-Assaad} An energy gap $\Delta _{\uparrow }$ exists in the spin-up
quasiparticle band, and there is a plateau in the total magnetization: $%
m_{c}+m_{f}=(1-n_{c})/2$.
\begin{figure}[tbp]
\includegraphics[width=1.8in,angle=-90]{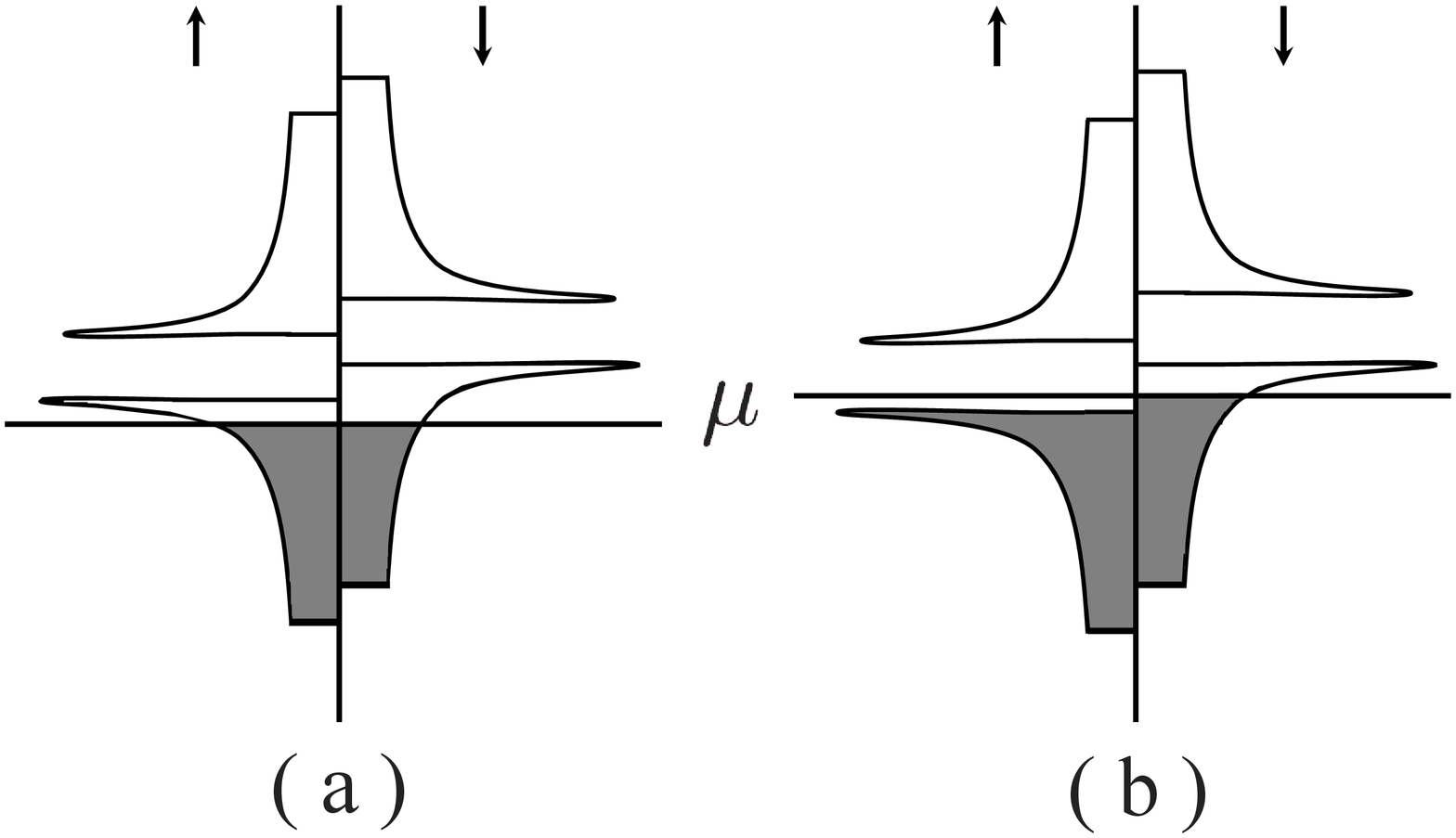}
\caption{Schematic DOS in the presence of Kondo screening effect.
(a) for the spin partially polarized ferromagnetic phase, (b) for
the spin fully polarized ferromagnetic phase.} \label{fig-DOS}
\end{figure}

The ground-state phase diagram has been obtained in our previous study.\cite%
{Li-Zhang-Yu} When $n_{c}<0.16$, the spin-polarized ferromagnetic phase is a
ground state in the large Kondo coupling region. For $0.16<n_{c}<0.82$, the
ground state is given by the spin partially polarized ferromagnetic phase in
the weak Kondo coupling limit; while in the intermediate Kondo coupling
regime, both spin fully polarized and partially polarized ferromagnetic
ordered phases with a finite value of the hybridization parameter $V$ may
appear, depending on the value of the Kondo coupling strength. For a strong
Kondo coupling, the pure Kondo paramagnetic phase is the ground state. There
is a continuous transition from the spin partially polarized ferromagnetic
ordered phase to the Kondo paramagnetic phase.

Now we calculate the finite temperature phase diagram. First of all, if the
temperature is high enough, all order parameters must disappear, so the
conduction electrons and local moments are decoupled. As the temperature is
decreased down to the Curie temperature of the pure ferromagnetic phase $%
T_{C}^{0}$, both $m_{c}$ and $m_{f}$ approach zero, but the ratio $%
m_{c}/m_{f}$ is finite. The self-consistent equations give rise to%
\begin{equation}
\lambda =\mu ,m_{c}\approx -\frac{J_{K}}{4D}m_{f},m_{f}=-\frac{J_{K}}{%
8T_{C}^{0}}m_{c},  \label{mc-mf-1}
\end{equation}%
and the Curie temperature $T_{C}^{0}$ can be estimated as
\begin{equation}
T_{C}^{0}=\frac{J_{K}^{2}}{32D},
\end{equation}%
which is independent of the density of conduction electrons, similar to the
characteristic energy scale given by the RKKY interaction.

On the other hand, if $J_{K}$ is large enough, the system must be in the
Kondo paramagnetic phase. As the Kondo coupling decreases, the Kondo
singlets are destabilized. When $T\rightarrow T_{K}^{0}$, the hybridization
vanishes, and the self-consistent equations can be reduced to
\begin{eqnarray}
\frac{1}{D}\int\limits_{-D}^{D}\frac{d\omega }{e^{(\omega -\mu )/T_{K}^{0}}+1%
} &=&n_{c},  \notag \\
\frac{J_{K}}{2D}\int\limits_{-D}^{D}d\omega \frac{\tanh (\frac{\omega -\mu }{%
T_{K}^{0}})}{\omega -\mu } &=&1.
\end{eqnarray}%
When we numerically solve these two equations, the Kondo temperature in the
paramagnetic phase $T_{K}^{0}$ can be obtained, which is the same
characteristic energy scale as derived from the $1/N$ expansion.\cite%
{read,coleman,burdin}

After obtaining $T_{C}^{0}$ and $T_{K}^{0}$, we expect that the pure
ferromagnetic phase exists for $T_{C}^{0}>T_{K}^{0}$ in the small Kondo
coupling limit, while for $T_{K}^{0}>T_{C}^{0}$ the Kondo screening is
present, and the competition between the ferromagnetic correlations and
Kondo screening effect should be taken into account more carefully.

In the presence of the Kondo screening, the corresponding Curie temperature $%
T_{C}$ can be still defined. As $T\rightarrow T_{C}$, the magnetic moments $%
m_{c}$ and $m_{f}$ approach zero, but their ratio is finite, $%
m_{c}/m_{f}\neq 0$. The self-consistent equations Eq.(5) can be solved
numerically, leading to the Curie temperature $T_{C}$ and the mean-field
parameters of $\mu $, $\lambda $, $V$, and $m_{c}/m_{f}$. On the other hand,
when the ferromagnetism is present, we can also introduce the Kondo
temperature $T_{K} $ incorporating the hybridization effect. When $%
V\rightarrow 0$ and $T\rightarrow T_{K}$, the numerical solution of the
self-consistent equations gives rise to the Kondo temperature $T_{K}$ and
the mean-field parameters $\mu $, $\lambda $, $m_{c}$, and $m_{f}$.

The resulting phase diagram is shown in Fig.3 for $n_{c}=0.2$. As the Kondo
coupling $J_{K}$ increases from a small value, the Curie temperature $T_{C}$
first increases up to a maximal value, and then continuously decreases to
zero at $J_{K}^{c2}=1.133D$. For small values of $J_{K}/D<0.41$, the Kondo
temperature $T_{K}$ vanishes. However, when $J_{K}/D>0.41$, the Kondo
temperature curve consists of two parts, meeting each other precisely at the
Curie temperature ($T_{K}=T_{C}$). Inside the ferromagnetic ordered phase, $%
T_{K}$ starts from a finite value and then decreases down to zero at $%
J_{K}^{c1}=0.506D$; while in the paramagnetic phase, $T_{K}$ follows the
behavior of the bare Kondo temperature $T_{K}^{0}$.
\begin{figure}[tbp]
\includegraphics[scale=0.5]{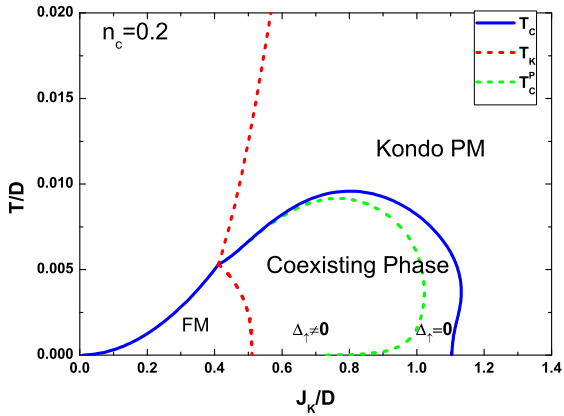}
\caption{(Color online) The finite temperature phase diagram at $n_{c}=0.2$.
In addition to the pure ferromagnetic ordered phase ($V=0$, $m_{c}\neq 0$
and $m_{f}\neq 0$) and the Kondo paramagnetic phase ($V\neq 0$, $%
m_{c}=m_{f}=0$), there are two different ferromagnetic ordered phases
coexisting with the Kondo screening ($V\neq 0$, $m_{c}\neq 0$ and $%
m_{f}\neq 0$): the spin fully polarized phase ($\Delta _{\uparrow
}\neq 0$) and spin partially polarized phase ($\Delta _{\uparrow
}=0$). The boundary of the pure ferromagnetic order phase and the
coexisting ferromagnetic ordered phases actually corresponds to a
crossover not a phase transition.}
\end{figure}

In the coexistence region, depending on whether the chemical potential $\mu $
is inside the energy gap of spin-up quasi-particle (as shown in Fig.2), we
can calculate the lowest excitation energy defined by $\Delta _{\uparrow
}\equiv \mu -\omega _{2+}$. In Fig.4, we show $\Delta _{\uparrow }$ as a
function of $T$ with fixed Kondo coupling parameters $J_{K}/D=0.6$, $0.7$, $%
0.733$, $0.9$ and $1.0$, respectively. It is clearly demonstrated
that the gap $\Delta _{\uparrow }$ has a non-monotonic behavior as
the temperature increases. Notice that $J_{K}/D=0.733$ corresponds
to the critical value between the spin fully polarized phase and
partially polarized phase at zero temperature. When $\Delta
_{\uparrow }\rightarrow 0$, the characteristic temperature
$T_{C}^{P}$ is determined, leading to the phase boundary
separating the spin fully polarized and the partially polarized
ferromagnetic ordered phases. The spin fully polarized
ferromagnetic order phase spans a large area in the coexistence
region, while the spin partially polarized phase just sits in a
narrow strip close to the phase boundary of the paramagnetic heavy
Fermi liquid phase. The existence of the partially polarized
ferromagnetic order phase can be expected before the system enters
into the paramagnetic metallic phase.
\begin{figure}[tbp]
\includegraphics[scale=0.5]{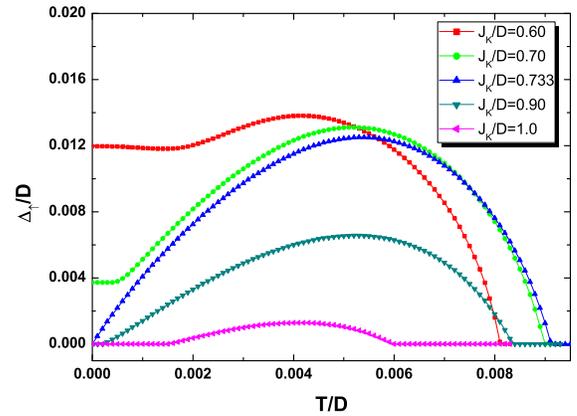}
\caption{(Color online) The energy gap of spin-up quasiparticles $\Delta
_{\uparrow }$ as a function of temperature $T$ in the coexisting phase at $%
n_{c}=0.2$.}
\end{figure}

Moreover, the magnetizations of the local moments and the
conduction electrons $m_{f}$ and $m_{c}$ are calculated as a
function of the Kondo coupling strength $J_{K}$ for $T=0.0025D$
and $0.0075D$, as shown in Fig.5., respectively. It is clear that
$m_{c}$ has an opposite sign of $m_{f}$, due to the
antiferromagnetic coupling between the local moments and
conduction electrons. In order to display the Kondo screening
effect, we have also plotted the hybridization parameter $V$ as a
function of $J_{K}$ for the same fixed temperatures. For low
temperatures shown in Fig.5a, the Kondo screening effect emerges
inside the ferromagnetic ordered phase, and a small drop is
induced in both magnetizations $m_{f}$ and $m_{c}$. When the
magnetization vanishes, the hybridization $V$ has a cusp. However,
for high temperatures shown in Fig.5b, the ferromagnetic ordering
appears inside the Kondo screened region. The cusps in the
hybridization curve are induced when the ferromagnetic order
parameters start to emergence or vanish.
\begin{figure}[tbp]
\includegraphics[scale=0.5]{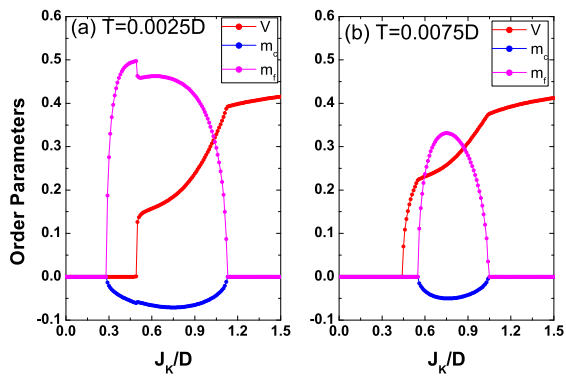}
\caption{(Color online) The ferromagnetic magnetizations and hybridization
parameters as a function of the Kondo coupling $J_{K}$ with a fixed
temperature at $n_{c}=0.2$. (a) $T=0.0025D$, (b) $T=0.0075D$.}
\end{figure}
\begin{figure}[tbp]
\includegraphics[scale=0.5]{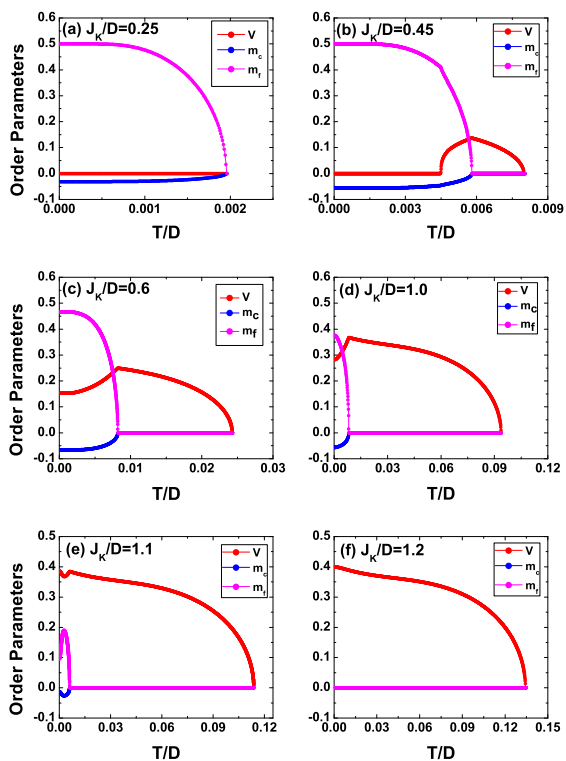}
\caption{(Color online) The magnetizations and hybridization parameter as a
function of temperature for a given Kondo coupling strength at $n_{c}=0.2$.
(a), (b), (c), (d), (e), and (f) correspond to $J_{K}/D=0.25$, $0.45$, $0.60$%
, $1.0$, $1.1$, and $1.2$, respectively. }
\end{figure}

The magnetizations $m_{c}$ and $m_{f}$ and hybridization parameter $V$ have
also been calculated as a function of temperature $T$ for the fixed Kondo
coupling strength $J_{K}$, which is shown in Fig.6. For a small value of $%
J_{K}/D=0.2$, as the temperature is increased, the magnetic moments $m_{c}$
and $m_{f}$ in the absence of the Kondo screening decrease down to zero at
the Curie temperature $T_{C}$ (see Fig.6a). In contrast, for a large value
of $J_{K}/D=1.2$, the system is in a paramagnetic heavy Fermi liquid phase
without ferromagnetic order (see Fig.6f).

For $J_{K}/D=0.45$, the Kondo screening effect starts to appear in the
presence of ferromagnetic ordering. When the ferromagnetic order disappears
at $T_{C}$, the hybridization reaches a maximal value and then decreases
down to zero at $T_{K}$ (see Fig.6b). For larger values of the Kondo
coupling $J_{K}/D=0.6$, $1.0$, and $1.1$, the Kondo screening effect
dominates in the temperature range, and the ferromagnetic ordering phase
occurs only in a small region, displayed in Fig.6c, Fig.6d and Fig.6e,
respectively. These three figures demonstrate the interplay between the
Kondo screening effect and the ferromagnetic correlations in the presence of
thermal fluctuations.

It is important to emphasize that all the results are obtained within the
effective mean field theory. When the fluctuation effects are incorporated
properly beyond the mean field level, the above phase transitions related to
the Kondo screening effect will be changed into a \textit{crossover}. Since
the Kondo screening order parameter, i.e. the effective hybridization is not
associated with a static long-range order, the finite $V$ does not
correspond to any spontaneous symmetry breaking. Therefore, in the obtained
finite temperature phase diagram Fig.3, only the Curie temperature $T_{C}$
(the solid line) represents a true phase transition.

Finally, it is important to mention a new Kondo lattice system YbNi$_{4}$P$%
_{2}$, recently discovered by distinct anomalies in susceptibility, specific
heat and resistivity measurements.\cite{Krellner-2011} Growing out of a
strongly correlated Kramers doublet ground state with Kondo temperature $%
T_{K}\sim 8K$, the ferromagnetic ordering temperature is severely reduced to
$T_{c}=0.17$K with a small magnetic moment $m_{f}\sim 0.05\mu _{B}$. Here we
would like to explain the small ferromagnetically order moment and the
substantially reduced Curie temperature as originating from the presence of
the Kondo screening effect, see Fig.6c, Fig.6d, and Fig.6e. The experimental
results can certainly be understood in terms of our effective mean field
theory. The quantum critical behavior observed experimentally requires a
quantum critical point separating the ferromagnetic ordered phase from the
Kondo paramagnetic phase at zero temperature, which is also consistent with
our finite temperature phase diagram. The further detailed calculations
concerning with thermodynamic properties of the heavy fermion ferromagnetism
are left for our future research.

In summary, within an effective mean-field theory for small
conduction electron densities $0.16<n_c<0.82$, we have derived the
finite temperature phase diagram. Inside the ferromagnetic ordered
phase, a characteristic temperature scale has been found to signal
the Kondo screening effect for the first time. In additional to
the pure ferromagnetic phase, there are two distinct ferromagnetic
long-range ordered phases coexisting with the Kondo screening
effect: a spin fully polarized phase and a partially polarized
phase. A second-order phase transition and a quantum critical
point have been found to separate the spin partially polarized
ferromagnetic ordered phase and the paramagnetic heavy Fermi
liquid phase. To some extent, our mean field theory has captured
the main physics of the Kondo lattice systems, which provides an
alternative interpretation of weak ferromagnetism observed
experimentally.

The authors acknowledge the support from NSF-China.

{\it Note added}. The ferromagnetic quantum critical point in the
heavy fermion metal YbNi$_4$(P$_{0.92}$As$_{0.08}$)$_2$ has been
further confirmed \cite{Steppke} by precision low temperature
measurements: the Gruneisen ratio diverges upon cooling to $T=0K$.


\begin{thebibliography}{99}
\bibitem{Stewart-2001} G. R. Stewart, Rev. Mod. Phys. \textbf{73}, 797
(2001).

\bibitem{Lohneysen-2007} H. v. Lohneysen, A. Rosch, M. Vojta, and P.
W\"olfle, Rev. Mod. Phys. \textbf{79}, 1015 (2007).

\bibitem{Steglich-2010} Q. Si and F. Steglich, Science \textbf{329}, 1161
(2010).

\bibitem{Doniach1977} S. Doniach, Physica, B \& C \textbf{91}, 231 (1977).

\bibitem{Lacroix1979} C. Lacroix, and M. Cyrot, Phys. Rev. B \textbf{20},
1969 (1979).

\bibitem{Q-Si} Q. Si, Physica B \textbf{378}, 23 (2006); Phys. Status
Solidi, B \textbf{247}, 476 (2010).

\bibitem{Zhang2000a} G. M. Zhang and L. Yu, Phys. Rev. B \textbf{62}, 76
(2000).

\bibitem{Assaad2001} S. Capponi and F. F. Assaad, Phys. Rev. B \textbf{63},
155114 (2001).

\bibitem{Ogata2007} H. Watanabe and M. Ogata, Phys. Rev. Lett. \textbf{99},
136401 (2007).

\bibitem{Assaad-2008} L. C. Martin and F. F. Assaad, Phys. Rev. Lett.
\textbf{101}, 066404 (2008).

\bibitem{Bernal-1995} O. O. Bernal, D. E. MacLaughlin, H. G. Lukefahr, and
B. Andraka, Phys. Rev. Lett. \textbf{75}, 2023 (1995).

\bibitem{Huy-2007} N. T. Huy, \textit{et al.}, Phys. Rev. B \textbf{75},
212405 (2007).

\bibitem{Krellner-2011} C. Krellner, \textit{et al.}, New J. Phys. \textbf{13%
}, 103014 (2011).

\bibitem{Flouquet-2011} A. Fernandez-Panella, D. Braithwaite, B. Salce, G.
Lapertot, and J. Flouquet, Phys. Rev. B \textbf{84}, 134416 (2011).

\bibitem{Lausberg-2012} S. Lausberg, \textit{et al.}, arXiv:1210.1345.

\bibitem{IK-1991} V. Y. Irkhin and M. I. Katsnelson, Z. Phys. B \textbf{82},
77 (1991).

\bibitem{Sigrist-1992} M. Sigrist, K. Ueda, and H. Tsunetsugu, Phys. Rev. B
\textbf{46}, 175 (1992).

\bibitem{Li-1996} Z. Z. Li, M. Zhuang, and M. W. Xiao, J. Phys.: Condens.
Matter \textbf{8} 7941 (1996).

\bibitem{Si} S. J. Yamamoto and Q. Si, Proc. Nat. Acad. Sci. \textbf{107},
15704 (2010).

\bibitem{Li-Zhang-Yu} G. B. Li, G. M. Zhang, and L. Yu, Phys. Rev. B \textbf{%
81}, 094420 (2010).

\bibitem{Peters-2012} R. Peters, N. Kawakami, and T. Pruschke, Phys. Rev.
Lett. \textbf{108}, 086402 (2012); R. Peters and N. Kawakami, Phys. Rev. B
\textbf{86}, 165107 (2012).

\bibitem{read} N. Read and D. N. Newns, J. Phys. C \textbf{16}, 3273 (1983).

\bibitem{coleman} P. Coleman, Phys. Rev. B \textbf{29}, 3035 (1984).

\bibitem{burdin} S. Burdin, A. Georges, and D. R. Grempel, Phys. Rev. Lett.
\textbf{85}, 1048 (2000).

\bibitem{Fazekas1991} P. Fazekas and E. M\"{u}ller-Hartmann, Z. Phys. B
\textbf{85}, 285 (1991).

\bibitem{Beach-Assaad} K. S. D. Beach and F. F. Assaad, Phys. Rev. B \textbf{%
77}, 205123 (2008); S. Viola Kusminskiy, K. S. D. Beach, A. H. Castro Neto,
and D. K. Campbell, Phys. Rev. B \textbf{77}, 094419 (2008).

\bibitem{Steppke} A. Steppke, et al., Science \textbf{339}, 933
(2013).
\end{thebibliography}
\end{document}